\documentstyle[aps,prl,epsf,twocolumn,floats]{revtex}
\begin{document}
\draft

%2col
\twocolumn[\hsize\textwidth\columnwidth\hsize\csname @twocolumnfalse\endcsname
% start of wide text
%2col

\title{\bf Renormalization-group study of a magnetic impurity in a
Luttinger liquid}
\author{Avraham Schiller\cite{present} and Kevin Ingersent\cite{byline}}
\address{Department of Physics, University of Florida,
P.\ O.\ Box 18440, Gainesville, Florida 32611-8440}
\date{16 September 1996}
\maketitle

\begin{abstract}
A generalized Anderson model for a magnetic impurity in an interacting
one-dimensional electron gas is studied via a mapping onto a classical
Coulomb gas.
For weak potential scattering, the local-moment parameter regime expands
as repulsive bulk interactions become stronger, but the Kondo scale for the
quenching of the impurity moment varies nonmonotonically.
There also exist two regimes dominated by backward potential scattering:
one in which the impurity is nonmagnetic, and another in which an unquenched
local moment survives down to very low temperatures.
\end{abstract}


%2col
% end of wide text
]
\narrowtext
%2col

The problem of a magnetic impurity in a Luttinger liquid touches upon two
important issues:
the response of an interacting one-dimensional (1D) electron system to a
local dynamical perturbation;
and the effect of bulk interactions on the Kondo effect.
The observation that a single potential scatterer in a 1D system can
cause the conductance to vanish \cite{potl} has generated great interest.
While the physics of a static impurity is by now largely understood, the
behavior of a dynamical impurity --- a magnetic atom being a natural
example --- remains less clear.
Even in a noninteracting host, a magnetic impurity can produce interesting
many-body physics, in the form of the Kondo effect \cite{Kondo64}.
The combined treatment of bulk interactions and impurity-induced correlations,
which may be necessary, for instance, to understand the heavy-fermion behavior
observed in Nd$_{2-x}$Ce$_x$CuO$_4$ \cite{ndcecuo4}, represents a considerable
theoretical challenge.

Most work on magnetic impurities in 1D systems \cite{1D-kondo} has focused
on the Kondo model, presupposing the existence of a local moment.
It has been found that repulsive bulk interactions within a Luttinger-liquid
host make possible a Kondo effect for ferromagnetic (as well as
antiferromagnetic) exchange couplings $J$, and also drive the form of the
Kondo temperature $T_K$ (below which the exchange renormalizes to strong
coupling) from an exponential in $J$ to a power-law.
Analysis of an Anderson impurity in a restricted Luttinger model
suggests that interactions expand the local-moment regime \cite{Phillips96}.
For the full Anderson problem, it has been predicted \cite{Li95} that a
single impurity modifies the bulk interactions, a conclusion that is not
supported by the present work.

In this paper we study a generalized Anderson model which includes the
direct and exchange interaction between localized and delocalized
electrons, in addition to hybridization between the impurity and
the Luttinger liquid.
This enables the first investigation of the competition between local-moment
formation (and a subsequent Kondo effect) and the correlated behaviors which
can arise due to strong backward potential scattering \cite{potl}.
We map the dynamical impurity problem onto a classical Coulomb gas, which
is analyzed using perturbative renormalization-group (RG) techniques.
We find that repulsive bulk interactions tend to stabilize local-moment
formation by suppressing charge transfer between the impurity and the host.
With increasing interactions, the Kondo temperature first rises,
as found in the pure Kondo model, but then falls due to the suppression
of hybridization.
In addition to a magnetic phase dominated by exchange scattering and a
nonmagnetic phase driven by backward potential scattering, there exists a
novel regime which exhibits a Curie susceptibility down to low temperatures,
but which displays the anomalous transport properties of a static impurity.

We begin with an extended Anderson Hamiltonian,
$
   H = H_{\text{\em Lutt}} + H_{\text{\em imp}} + H_{\text{\em hybrid}} +
       H_{\text{\em Kondo}} + H_{density}.
$
Here,
\begin{equation}
   H_{\text{\em Lutt}} =
      \sum_{\nu = \rho, \sigma}
      \frac{v_{\nu}}{2}\int \left[ K_{\nu}(\nabla \theta_{\nu})^2 +
      K_{\nu}^{-1} (\nabla \phi_{\nu})^2 \right ] dx
                                                        \label{H_Lutt}
\end{equation}
is the Luttinger-liquid Hamiltonian for a 1D system,
written in terms of four bosonic fields,
$\theta_{\nu}(x)$ and $\phi_{\nu}(x)$, which describe independent charge
($\nu=\rho$) and spin ($\nu=\sigma$) density modes \cite{1D}.
These fields are related through their gradients to the local charge- and
spin-density operators for the left- and right-moving fermion
fields, $\Psi_{-,s}(x)$ and $\Psi_{+,s}(x)$, respectively.
(We use $s=\uparrow,\downarrow$ and $+,-$ interchangeably to label the
spin $z$ component.)
The Luttinger liquid is characterized by four parameters:
the separate sound velocities $v_{\nu}$ and interaction parameters
$K_{\nu}$ for charge and spin modes.
In the absence of bulk interactions, $K_{\rho} = K_{\sigma} = 1$.
Repulsive bulk interactions push down $K_{\rho}$ into the range
$0 < K_{\rho} < 1$, but SU(2) spin symmetry requires that $K_{\sigma}$
remains equal to unity.

The nondegenerate impurity is described in terms of its energy $\varepsilon_d$
and the on-site repulsion $U$ between a pair of localized $d$ electrons:
\begin{equation}
   H_{\text{\em imp}} = \varepsilon_d n_d + \frac{U}{2} n_d (n_d-1) ,
   \quad n_d \in \{0,1,2\} .
                                                        \label{H_imp}
\end{equation}

The remaining terms in $H$ contain the interaction between the orbitals
$d_s$ and the fermion fields $\Psi_{p,s}(0)$ at the impurity site ($x = 0$):
the hybridization,
\begin{equation}
   H_{\text{\em hybrid}} =
      t\sqrt{a} \sum_{p = \pm} \sum_{s = \uparrow, \downarrow}
      \left\{ d^{\dagger}_{s}\Psi_{p,s}(0) +
      \Psi^{\dagger}_{p,s}(0) d_{s} \right\} ;
                                                        \label{H_t}
\end{equation}
the direct spin-exchange (Kondo) interaction,
\begin{eqnarray}
   H_{\text{\em Kondo}} &=&
      \frac{J_{z F}a}{4}\sum_{p = \pm} \sum_{s, s'} s s'
      d^{\dagger}_{s'}d_{s'}
      \Psi^{\dagger}_{p,s}(0) \Psi_{p,s}(0) \nonumber\\
  &+& \frac{J_{\perp F} a}{2}\sum_{p = \pm}
      \left\{ d^{\dagger}_{\uparrow}d_{\downarrow}
      \Psi^{\dagger}_{p,\downarrow}(0)
      \Psi_{p,\uparrow}(0) + \text{H.c.} \right \} \nonumber\\
  &+& \frac{J_{z B} a}{4}\sum_{p = \pm} \sum_{s, s'} s s'
      d^{\dagger}_{s'}d_{s'}
      \Psi^{\dagger}_{p,s}(0) \Psi_{-p,s}(0) \nonumber\\
  &+& \frac{J_{\perp B} a}{2}\sum_{p = \pm}
      \left\{ d^{\dagger}_{\uparrow}d_{\downarrow}
      \Psi^{\dagger}_{p,\downarrow}(0)
      \Psi_{-p,\uparrow}(0) + \text{H.c.} \right \};
                                                        \label{H_Kondo}
\end{eqnarray}
and the density-density interaction,
\begin{eqnarray}
   H_{\text{\em density}} &=&
      V_F a \sum_{p = \pm}\sum_{s = \uparrow, \downarrow}
      n_d \; :\!\Psi^{\dagger}_{p,s}(0) \Psi_{p,s}(0)\!: \nonumber\\
  &+& V_B a \sum_{p = \pm} \sum_{s = \uparrow, \downarrow}
      n_d \; \Psi^{\dagger}_{p,s}(0) \Psi_{-p,s}(0).
                                                        \label{H_density}
\end{eqnarray}
In the above equations, $a$ is a lattice spacing;
$t$ is the hybridization matrix element;
$J_{z F}$ and $J_{\perp F}$ ($J_{z B}$ and $J_{\perp B}$) are the
longitudinal and transverse couplings for forward (backward)
exchange scattering;
and $V_F$ ($V_B$) is the forward (backward) potential scattering.
Note that SU(2) spin symmetry requires that
$J_{z F} = J_{\perp F}$ and $J_{z B} = J_{\perp B}$.
The local density $\Psi^{\dagger}_{p,s}(0) \Psi_{p,s}(0)$
in Eq.~(\ref{H_density}) is normal ordered with respect
to the unperturbed Fermi sea.

We study this model by extending Si and Kotliar's treatment of a
generalized Anderson impurity in a Fermi liquid \cite{Si93}, which
combines Haldane's mapping of the Anderson model onto a classical
Coulomb gas \cite{Haldane78} with an RG procedure due to Cardy \cite{Cardy81}.
Below we sketch the main steps of our calculation, which has the virtue
that it is nonperturbative in the bulk interactions.
A detailed account will be given elsewhere.

We separate the Hamiltonian into an unperturbed part $H_0$, plus a
perturbation $H_1$ which contains all terms that change the number of
electrons in any of the four conduction-electron branches
($+,\!\uparrow; +,\!\downarrow; -,\!\uparrow$; and $-,\!\downarrow$).
Thus, $H_0$ comprises $H_{\text{\em Lutt}}$, the $J_{z F}$ Kondo terms,
and the $V_F$ potential-scattering terms.
We expand the partition function $Z$ in powers of $H_1$, and express it as
a sum over all possible histories.
A history is a sequence of ``kinks'' --- each kink corresponding to one
application of a term in $H_1$ --- which overall preserves the impurity
configuration and the occupation of each conduction branch.
The existence of both left- and right-moving fermions requires that we
follow the history not only of the impurity configuration, but also of the
branch occupation numbers.

\begin{table}[t]
\renewcommand{\arraystretch}{1.22}
\begin{tabular}{clc}
$A$ & $dA/d\ln(D_0/D) =$ &
   $A_{\text{\em bare}}$ \\[0.3ex]
\tableline
$y_{t}$ &
   $(1\!-\!k_t) y_{t} + y_{t} (y_{1 V} + \frac{1}{2}y_{z B} +
   y_{\perp F}$ &
      $t / \sqrt{2}D_0$ \\
   &
   $\qquad + y_{\perp B}) e^{-E/2} - y_{t} y_{0 V}e^{E/2}$ &
      \\
$y_{\perp F}$ &
   $(1\!-\!k_{\perp F}) y_{\perp F} + 2y_{\perp B} y_{z B} + y_{t}^2 e^E$ &
      $J_{\perp F} / 4D_0$ \\
$y_{\perp B}$ &
   $(1\!-\!k_{\perp B}) y_{\perp B} + 2y_{\perp F} y_{z B} + y_{t}^2 e^E$ &
      $J_{\perp B} / 4D_0$ \\
   &
   $\qquad + 2y_{\perp B} y_{1P}$ &
      \\
$y_{z B}$ &
   $(1\!-\!k_{z B}) y_{z B} + 4 y_{\perp F}y_{\perp B} + y_{t}^2 e^E$ &
      $J_{z B} / 4D_0$ \\
   &
   $\qquad + 2y_{z B} y_{1P}$ &
      \\
$y_{1 V}$ &
   $(1\!-\!k_V) y_{1 V} + \frac{1}{2}y_{t}^2 e^E - 2y_{1 V} y_{1P}$ &
      $V_B / 2D_0$ \\
$y_{0 V}$ &
   $(1\!-\!k_V) y_{0 V} - y_{t}^2e^{-E} - 2y_{0 V} y_{0P}$ &
      $0$ \\
$y_{1P}$ &
   $(1\!-\!k_P) y_{1P} - 2y_{1 V}^2 + \frac{1}{2} y_{z B}^2 + y_{\perp B}^2$ &
      $0$ \\
$y_{0P}$ &
   $(1\!-\!k_P) y_{0P} - 2y_{0 V}^2$ &
      $0$ \\
$\epsilon_{\sigma}$ &
   $(1\!-\!2 \epsilon_{\sigma}) \left( y_{t}^{2}e^E +
   2y_{\perp F}^2 + 2y_{\perp B}^2 \right) $  &
      $J_{z F} / 4D_0$ \\
$\epsilon_{\rho}$ &
   $(1\!-\!2\epsilon_{\rho}) y_{t}^{2} \left( 2e^{-E} + e^E \right)$ &
      $K_{\rho} V_F / D_0$ \\
$E$ &
   $E + 4y_{0 V}^2 - 4y_{1 V}^2 - y_{z B}^2 - 2y_{\perp F}^2$ &
      $\varepsilon_d /D_0$ \\
   &
   $ \qquad - 2y_{\perp B}^2 + 2y_{t}^2 \left( 2e^{-E} - e^E \right)$ &
      $- \left( J_{z F} /4D_0 \right)^2 $ \\
   &
   $\qquad + 2y_{0P}^2 - 2y_{1P}^2$ &
      $- K_{\rho} \left( V_F /D_0 \right)^2 $
\end{tabular}
\vspace{2ex}
\caption{Summary of the kink-gas parameters $A$ for the infinite-$U$ case,
their bare values $A_{\text{\em bare}}$, and the corresponding RG equations.
Here $y_{\gamma}$ are fugacities of the kink-gas action,
$k_{\gamma}$ are their corresponding scaling dimensions, and $E$ is a
symmetry-breaking field. The coefficients $\epsilon_{\sigma}$
and $\epsilon_{\rho}$ enter the scaling dimensions, which have
the following explicit forms:
$k_t = [K_{\rho} + K_{\rho}^{-1}(1 - 2\epsilon_{\rho})^2
+ 1 + (1 - 2\epsilon_{\sigma})^2]/8$,
$k_{\perp F} = [1 + (1 - 2\epsilon_{\sigma})^2]/2$,
$k_{\perp B} = [K_{\rho} + (1 - 2\epsilon_{\sigma})^2]/2$,
$k_{z B} = k_V = (1 + K_{\rho})/2$,
and
$k_P = 2 K_{\rho}$.
}
\label{table:RG}
\end{table}

The response of the Luttinger liquid to local distortions gives rise to
long-range interactions between the kinks within a given history.
Hence, just as in a Fermi liquid host \cite{Si93}, $Z$ takes the form of
a fugacity expansion for the partition function of a 1D multicomponent gas
of kinks with logarithmic interactions.
In a Luttinger liquid, however, the number of fugacities (one for each
distinct type of kink) is much greater than in a Fermi liquid, and the
coupling between kinks depends explicitly on $K_{\rho}$.

The endpoint of these manipulations is a set of differential equations
describing the renormalization of dimensionless kink-gas parameters
with increasing short-time cutoff, the inverse of the bandwidth
$D=\pi v_{\nu}/a$.
(We assume equal velocities for the spin and charge sectors.)
Table~\ref{table:RG} lists these equations, restricted for brevity to
the case $U = \infty$.
The effect of bulk interactions is to change the scaling dimensions
which measure the relevance or irrelevance of the fugacities.
The first five fugacities in Table~\ref{table:RG} correspond to terms in the
original Hamiltonian.
Additional kinks are generated upon renormalization, among which the most
relevant are backward potential scattering from an empty impurity ($y_{0 V}$)
and singlet $4 k_F$ pair-scattering from an empty or singly occupied impurity
($y_{0P}$ or $y_{1P}$).
[The pair-scattering processes represent Hamiltonian terms proportional to
$\Psi^{\dagger}_{+,\uparrow}(0) \Psi^{\dagger}_{+,\downarrow}(0)
\Psi_{-,\downarrow}(0)\Psi_{-,\uparrow}(0) + \text{H.c.}$]
Table~\ref{table:RG} contains three other parameters:
$\epsilon_{\sigma}$ and $\epsilon_{\rho}$, which enter the kink-gas
formalism via the effective charges associated with the kinks, and correspond,
respectively, to $J_{z F}$ Kondo scattering and the
forward-potential-scattering difference between singly occupied and empty
impurity states;
and a symmetry-breaking field $E$, representing the energy difference
between the singly occupied and empty states.

\begin{figure}
\centerline{
\vbox{\epsfxsize=80mm \epsfbox{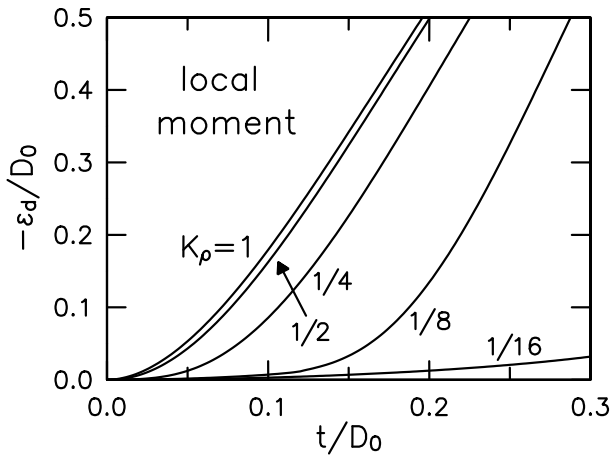}}
}
\vspace{2ex}
\caption{
Boundary of the local-moment regime, for $U=\infty$, zero bare exchange and
potential scattering, and different bulk interaction parameters $K_{\rho}$
(labeled).
The local-moment regime lies above and to the left of each curve.
}
\label{fig:local}
\end{figure}

By construction, the kink-gas approach treats $J_{z F}$ and $V_F$ on
a different footing from their backward-scattering and spin-flip
counterparts, leading to nonsystematic third-order (in $y$ and $\epsilon$)
terms in the RG equations.
Upon truncation to second order, though, our equations properly preserve SU(2)
spin symmetry, in contrast to those derived by Lee and Toner for the
Kondo model \cite{1D-kondo}.
We also note that there is no renormalization of the bulk parameter $K_{\rho}$,
contrary to Ref.~\onlinecite{Li95}.

Given bare values of the kink-gas parameters, the system of coupled RG
equations in Table~\ref{table:RG} can be iterated while the bandwidth $D$
is reduced from its bare value $D_0$.
This can continue until the magnitude of any parameter becomes of order unity,
at which point the fugacity expansion breaks down.
The low-temperature physics is determined by which particular parameter
becomes large first.
For instance, we identify the local-moment regime as that within which at
least one parameter from among $-E$, $|y_{\perp F}|$, $|y_{\perp B}|$,
$|y_{zB}|$, $|y_{1V}|$ or $|y_{1P}|$ (typically, it is $-E$) grows to $+1$,
while all others remain smaller in magnitude.
As shown in Fig.~\ref{fig:local}, the range of parameters $t$ and
$\varepsilon_d$ satisfying this criterion grows monotonically as $K_{\rho}$
decreases.
The expansion is initially slight, but becomes more dramatic once
$K_{\rho}$ falls below about $0.25$.

\begin{figure}
\centerline{
\vbox{\epsfxsize=80mm \epsfbox{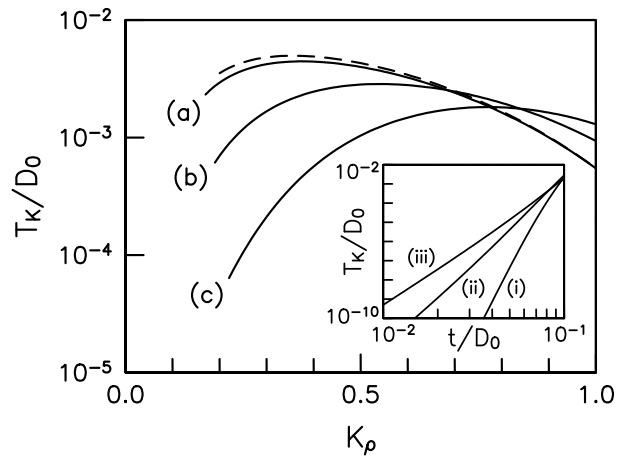}}
}
\vspace{2ex}
\caption{
Kondo temperature $T_K$ vs bulk interaction parameter $K_{\rho}$,
for zero bare exchange and potential scattering.
Solid curves: $U = \infty$ and
(a) $\varepsilon_d = -0.5D_0$, $t = 0.15D_0$;
(b) $\varepsilon_d = -0.24D_0$, $t = 0.10D_0$;
(c) $\varepsilon_d = -0.06D_0$, $t = 0.05D_0$.
Broken curve: $U/2 = -\varepsilon_d = 0.5D_0$, $t = 0.11D_0$.
Inset:
Log-log plot of $T_K$ vs hybridization $t$, for $U = \infty$,
$\varepsilon_d = -0.24D_0$, and
(i) $K_{\rho} \!=\! 3/4$;
(ii) $K_{\rho} \!=\! 1/2$;
(iii) $K_{\rho} \!=\! 1/3$.
}
\label{fig:Tk}
\end{figure}

The expansion of the local-moment regime stems from the reduction in the
scaling dimension of $y_t$ ($k_t$ in Table~\ref{table:RG}), the precise value
of which derives from the power-law density of states of a Luttinger liquid
\cite{1D}.
Physically, however, this trend simply reflects the increased difficulty of
transferring an electron between the impurity and a correlated host.
Similar effects have been found both in a restricted Luttinger model
\cite{Phillips96} and in a Fermi liquid with a power-law density of states
\cite{Buxton96}.
We expect the same to be true of other interacting systems, as well.

In a Fermi liquid, starting from weak coupling, the only alternatives to
local-moment formation are mixed valence ($y_t$ becomes large)
and an empty impurity ($E$ grows to $+1$).
A Luttinger liquid can support additional phases in which
more than one RG parameter is large (e.g., $y_t$ and $y_{0V}$).
Generalized mixed-valence behavior occurs even for bulk interactions so strong
($K_{\rho} < 3-\sqrt{8}\approx 0.17$) that $y_t$ is formally irrelevant
($k_t>1$).

If a local moment does form, around some energy scale $T_{\text{\em LM}}$,
the possibility arises of a Kondo effect for
$D\lesssim T_K < T_{\text{\em LM}}$.
Within the Kondo model, $T_K$ increases monotonically with bulk interactions
and, when $J\ll(1-K_{\rho})D_0$, crosses over from an exponential in $J$
(the form found in a Fermi liquid) to a power-law:
$T_K\approx$ $D_0[J/(1-K_{\rho})D_0]^{2/(1-K_{\rho})}$ \cite{1D-kondo}.
Since our RG equations extend to the local-moment regime (once the empty
impurity state is eliminated via a Schrieffer-Wolff transformation),
we can study the dependence of $T_K$ on the parameters of the more
fundamental Anderson model.

\begin{figure}
\centerline{
\vbox{\epsfxsize=80mm \epsfbox{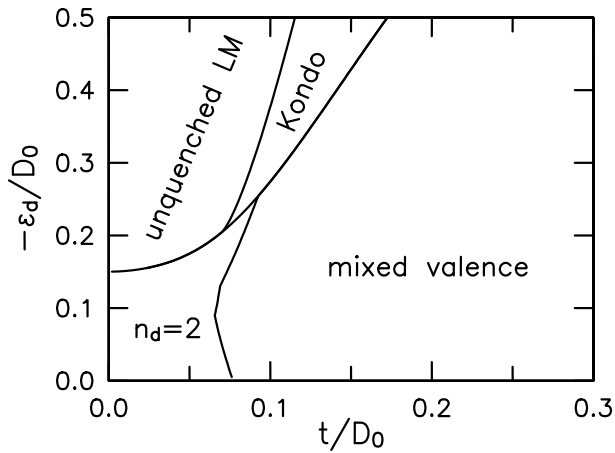}}
}
\vspace{2ex}
\caption{
Phase diagram, $\varepsilon_d$ vs\ $t$, for $K_{\rho}=1/3$ and a
symmetric Anderson impurity ($U=-2\varepsilon_d$) with $V_B = 0.1D_0$.
See the text for an explanation of the different phases.
}
\label{fig:phase}
\end{figure}

Figure~\ref{fig:Tk} shows the variation of $T_K$ with $K_{\rho}$ for
fixed $\varepsilon_d$ and $t$.
Qualitatively the same behavior is seen for $U=\infty$ and for
the opposite limit of a symmetric impurity, $U=-2\varepsilon_d$.
The nonmonotonic variation with decreasing $K_{\rho}$ reflects a competition
between (a) reduced hybridization at high energies,
$D\gtrsim T_{\text{\em LM}}$, which depresses the effective exchange
$J_{\text{\em LM}}$ on entry to the local-moment regime,
and (b) accelerated renormalization of the Kondo couplings for
$D\lesssim T_{\text{\em LM}}$.
For weak interactions ($K_{\rho}$ close to $1$), the latter effect wins,
and the variation of $T_K$ is similar to that found in the Kondo model
\cite{1D-kondo}.
For stronger interactions, though, the depression of $J_{\text{\em LM}}$
dominates: $T_K$ is still approximately proportional to $t^{4/(1-K_{\rho})}$
(see inset in Fig.~\ref{fig:Tk}) --- consistent with the pure-Kondo case
\cite{1D-kondo} --- but nonuniversal prefactors reverse the overall trend with
$K_{\rho}$.
The position of the maximum in $T_K$ depends on the relative ranges of $D$
over which the system remains in the high-energy and local-moment regimes.
For $K_{\rho}\lesssim 0.2$, the local-moment regime can still be entered, but
subsequently the singlet pair-scattering $y_{1P}$ becomes comparable to
$y_{z B}$ and $y_{\perp B}$, making it unclear whether or not a Kondo effect
takes place.

To this point we have considered only cases with $V_F=V_B=0$.
In a Fermi liquid, potential scattering is exactly marginal and can generally
be absorbed into a redefinition of the bare impurity parameters.
In a Luttinger liquid, by contrast, backward potential scattering is
relevant \cite{potl}, and has the same scaling dimension as backward Kondo
scattering.
There is thus a delicate balance between magnetic and nonmagnetic scattering.

Figure~\ref{fig:phase} shows the $\varepsilon_d$ vs $t$ phase diagram for a
symmetric impurity ($U=-2\varepsilon_d$) with $V_F = V_B = 0.1 D_0$.
The boundaries mark fairly rapid crossovers between different regimes.
In addition to the Kondo and generalized mixed valence phases
described above, there are two regimes dominated by potential scattering.
The first occurs when $|\varepsilon_d|$ and $t$ are
sufficiently small that the backward potential scattering is able to
freeze the impurity in a doubly occupied ($n_d = 2$) configuration.
The problem then becomes equivalent to strong static potential scattering,
i.e., the electrical conductivity should vanish with temperature $T$ as
$T^{1/K_{\rho}-1}$ \cite{potl}.
This regime, in which the impurity is nonmagnetic, grows upward and to the
right in Fig.~\ref{fig:phase} as $V_B$ increases  or $K_{\rho}$ decreases.

The second regime cannot possibly develop in a Fermi liquid, starting from
weak coupling.
A local moment is formed, but subsequently the backward potential
scattering from the singly occupied impurity ($y_{1V}$ and $y_{1P}$)
becomes strong before the Kondo coupling can quench the local moment.
Over an extended temperature range the conductivity should again
vary as $T^{1/K_{\rho}-1}$, but in this regime the susceptibility
should be Curie-like.
(At very low energies, other screening processes may
eventually quench the impurity moment \cite{Fabrizio95}.)
As the bulk interactions become stronger, this unquenched region occupies
an increasing fraction of the local-moment regime.

In this paper, we have studied the competition between magnetic and
nonmagnetic scattering from an impurity atom placed in an interacting
1D electron gas.
The bulk correlations hinder charge transfer to and from the impurity,
thereby tending to stabilize the local-moment regime against mixed-valence
behavior.
However, strong backward potential scattering inhibits local-moment formation.
Bulk interactions also introduce a novel phase in which a local moment forms,
but the subsequent growth of backward potential scattering overwhelms the
Kondo effect, leaving an unquenched magnetic moment down to very low
energies or temperatures. In regions where a Kondo effect does take place,
$T_K$ varies nonmonotonically with increasing bulk interactions.

Portions of this work were carried out while one of us (K.\ I.) was visiting
the Institute for Theoretical Physics, Santa Barbara.
This research was supported in part by the NSF through the National
High Magnetic Field Laboratory and Grants DMR93--16587
and PHY94--07194.

\end{document}